# Individual and Group Dynamics in Purchasing Activity

Lei Gao Jin-Li Guo* Chao Fan Xue-Jiao Liu

*Business School, University of Shanghai for Science and Technology, Shanghai, 200093, China*

**Abstract**: As a major part of the daily operation in an enterprise, purchasing frequency is of constant change. Recent approaches on the human dynamics can provide some new insights into the economic behaviors of companies in the supply chain. This paper captures the attributes of creation times of purchase orders to an individual vendor, as well as to all vendors, and further investigates whether they have some kind of dynamics by applying logarithmic binning to the construction of distribution plot. It's found that the former displays a power-law distribution with approximate exponent 2.0, while the latter is fitted by a mixture distribution with both power-law and exponential characteristics. Obviously, two distinctive characteristics are presented for the interval time distribution from the perspective of individual dynamics and group dynamics. Actually, this mixing feature can be attributed to the fitting deviations as they are negligible for individual dynamics, but those of different vendors are cumulated and then lead to an exponential factor for group dynamics. To better describe the mechanism generating the heterogeneity of purchase order assignment process from the objective company to all its vendors, a model driven by product life cycle is introduced, and then the analytical distribution and the simulation result are obtained, which are in good line with the empirical data.

## 1 Introduction

Humans participate in a large number of distinct activities on different purposes and these activity patterns show diverse characteristics. Although individual human behaviors are unpredictable, while all the behaviors reveal some general laws. However, it is impossible to describe all the activity patterns with a specific process model. In 1837, French mathematician Poisson derived the Poisson process based on the binomial distribution and from then on, similar models are widely used in communications, transportation, management, economic operation and so on. All these models are based on the hypothesis that the time intervals between two consecutive actions are comparable to each other. In 2005, Barabási et al. studied the e-mail communications and the correspondence patterns of Einstein et al. and indicated that human activity patterns are rather heterogeneous, with long periods of inactivity that separate bursts of intensive activity. More precisely, the inter-event times between two consecutive executions for a given task can be approximated by a heavy tailed distribution [1-3]. After that, some scholars conducted related empirical exploration and theoretical research through human daily behaviors and found that power-law distributions occur in an extraordinarily diverse range of activities such as web browsing, communications via electronic and paper mails and human traffic [4-7]. According to the present research, the scaling parameter $\alpha$ typically lies in the range $1 < \alpha < 3$, although there are occasional exceptions. Later, Vázquez et al. divided human dynamics into two universality classes with exponents 1 and 1.5 [8]. The fact that these patterns can't be characterized simply is a sign of complex underlying human dynamics that merit further study.

* Corresponding author. Tel.:021-65710380.
*E-mail address*: phd5816@163.com (J.-L. Guo).

However, studies on the properties and mechanisms of human business activities based on the enterprise, an important organization in the real world, are still in the infancy, with only a few related documents. Wang et al. studied the statistical features of the task-restricted work patterns via aerial inbound operation in a logistics company and obtained five power-law distributions with exponents 1.5, 2.0 and 2.5[9]. Caldarelli et al. illustrated that some financial systems display scale-free properties, with exponents from 1.5 to 3[10]. Ho et al. investigated the time series of Taiwan stock price index and got a conclusion that the autocorrelation does not decay to zero exponentially but in a power-law manner with exponent $5/3$ over a range of frequencies [11]. Hence, it still needs further study to clarify whether the day-to-day business in a company, for example, supply chain operation, production scheduling and staff turnover, has heavy-tailed property or follows Poisson process.

Among all the daily activities in a manufacturing enterprise, supply chain management plays an important role. At present, in the dynamic environment and uncertain competitive market, the value and service provided by a good supply chain alliance is more competitive and effective than that by companies without information sharing between each other. Yet, as a separate individual in the supply chain level, a company possesses its unique view and strategy, forming a complicated relationship with other corporations under the effect of the environment and showing a power-law degree distribution with exponent 1.96[12]. And the purchase order to a large extent is an epitome of the games of all parties in the complex system. Recent approaches on the human dynamics can provide some new insights into the purchasing behaviors in the supply chain based on the time series of purchase orders and in turn, the results may give a further understanding of the characteristic of human activity pattern. The goal of this paper is to provide an empirical evidence and theoretical analysis for the purchasing process.

## 2 Data specification

In this paper, the data come from a main business of a Fortune 500 company, called Enterprise M for short in the following. As a corporation with global coverage, the number of daily purchase orders (PO's) for this business is as large as one thousand. Under the premise of keeping the business information of enterprise M absolutely secret, we investigate the PO's created in the two latest fiscal quarters and analyze the scaling characteristics with these actual operation data from the angle of time intervals. There are about $4\times10^4 \sim 5\times10^4$ PO's in each quarter, involving approximately two thousand suppliers. This paper focuses on the inter-arrival time of two concessive orders, with the precision of a second. But for the global property of enterprise M and the actual state in daily work, we don't classify the orders by the creation time, i.e., orders placed in the weekends and other nonworking times of a certain region are not precluded.

The analysis of each time interval between two consecutive purchase orders is performed by the normalized logarithmic binning to remove the noise tail and reduce the uneven statistical fluctuations common in empirical analysis. In contrast with simple logarithmic binning, it further removes the artifact while maintaining the advantage of using larger bins where there are fewer values of abscissa [13]. Naturally, prior to conducting the analysis, we compared the three fitting curves generated by this approach, probability density curves and cumulative distribution curves on logarithmic scales respectively, and finally decided to apply this method to data analysis for it is more explicit and legible on the premise of fitting accuracy without altering the real distribution characteristics of the orders [14].

During the following section, we mainly discussed two types of distributions on group and individual dynamics levels respectively: one is the time interval distribution of all the purchase orders of enterprise M without recognition of the PO suppliers, the other is of orders to a specific supplier.

**Table 1 Data Profile**

| Fiscal Quarter | Q1 | Q2 |
|---|---|---|
| Number of all PO's | $4.4\times10^4$ | $4.2\times10^4$ |
| Number of vendors | $2.7\times10^3$ | $2.6\times10^3$ |
| Number of PO's to the top ten vendors | $2.4\times10^4$ | $2.4\times10^4$ |
| Number of PO's to Enterprise A | $1.0\times10^4$ | $1.1\times10^4$ |

As shown in table 1, enterprise M placed PO's to more than two thousand vendors during the two quarters, among which about fifty percent were assigned to only ten suppliers and over twenty percent to enterprise A. In the case of the heterogeneity during the PO assignment, here we just discuss the purchase orders to the largest supplier A in details as a paradigm to clarify the individual dynamics of the purchasing process.

# 3 Individual dynamics in the purchasing activity

The individual characteristic of purchasing activity is analyzed through the statistical properties of the time intervals of orders to a specific vendor. Obviously, the individual purchasing frequency may show great variety with time. As demonstrated in Fig.1 and Fig.2, the PO's from enterprise M to its supplier A show a power law shape over a range of interval times, which reveals the heterogeneity of purchasing activities.

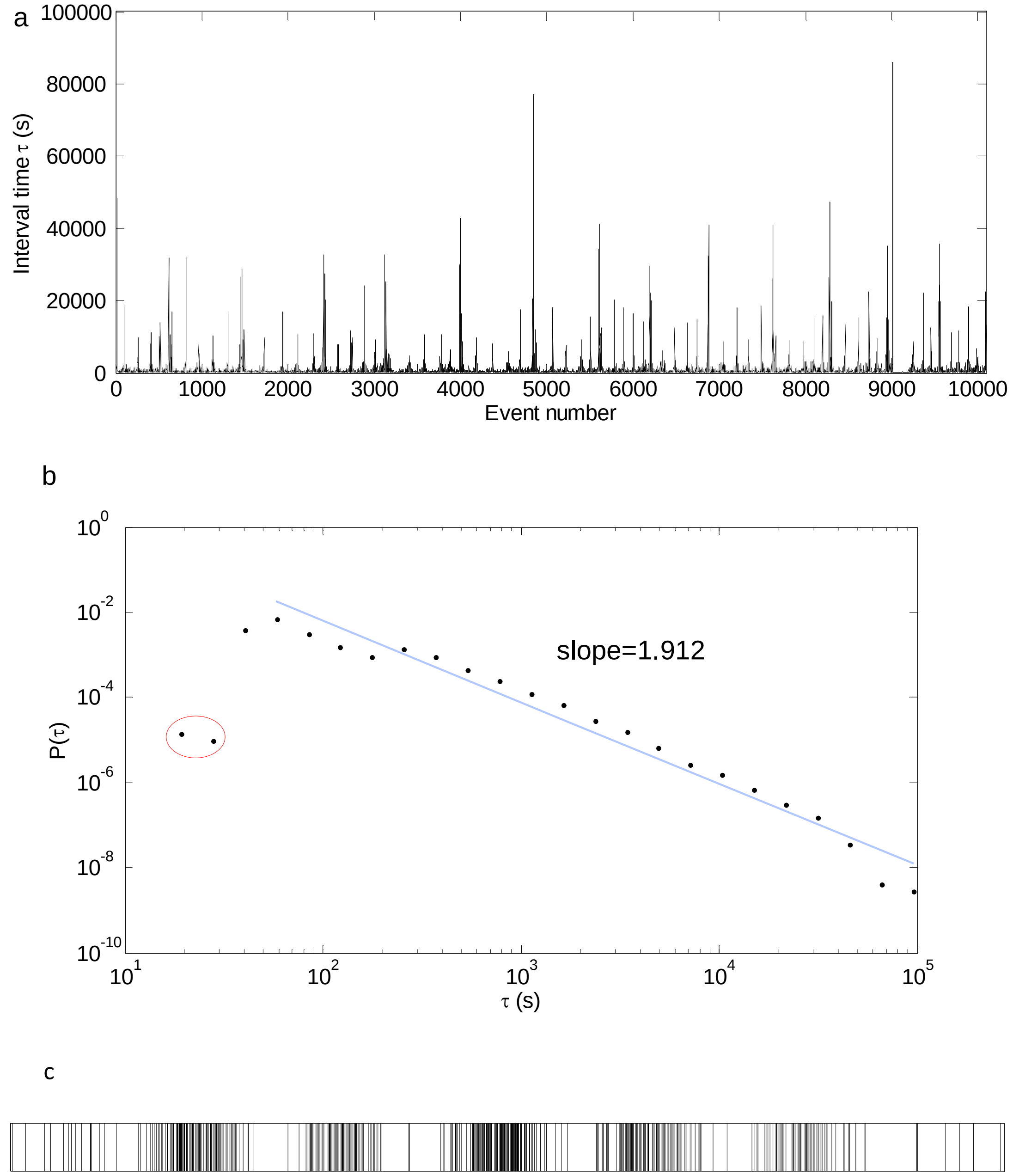

Fig.1 Purchase orders to the largest supplier A in Q1. **a**, The interval times $\tau$ of the $1.0\times10^{4}$ consecutive PO's to enterprise A in Q1, locating in range between 16s and $8.6\times10^{4}$ s. There are dozens of large spikes in this plot, implying very long interval times for orders created successively. **b**, Probability-interval time plot for these PO's in a log-log coordinate system, where time interval distribution $P(\tau)\propto\tau^{-1.912}$ is represented by a solid line linear. While conducting the regression analysis, the first two points marked in red are excluded for the obvious deviation. **c**, Succession of PO's in a certain week of Q1, from Sunday to Saturday. The horizontal axis denotes time with unit seconds, and each corresponding vertical line presents the creation of a PO to enterprise A. Five significant bursts (black boxes) are visible on the plot, the spacing of each vertical line illustrating the gaps seen in a. b.

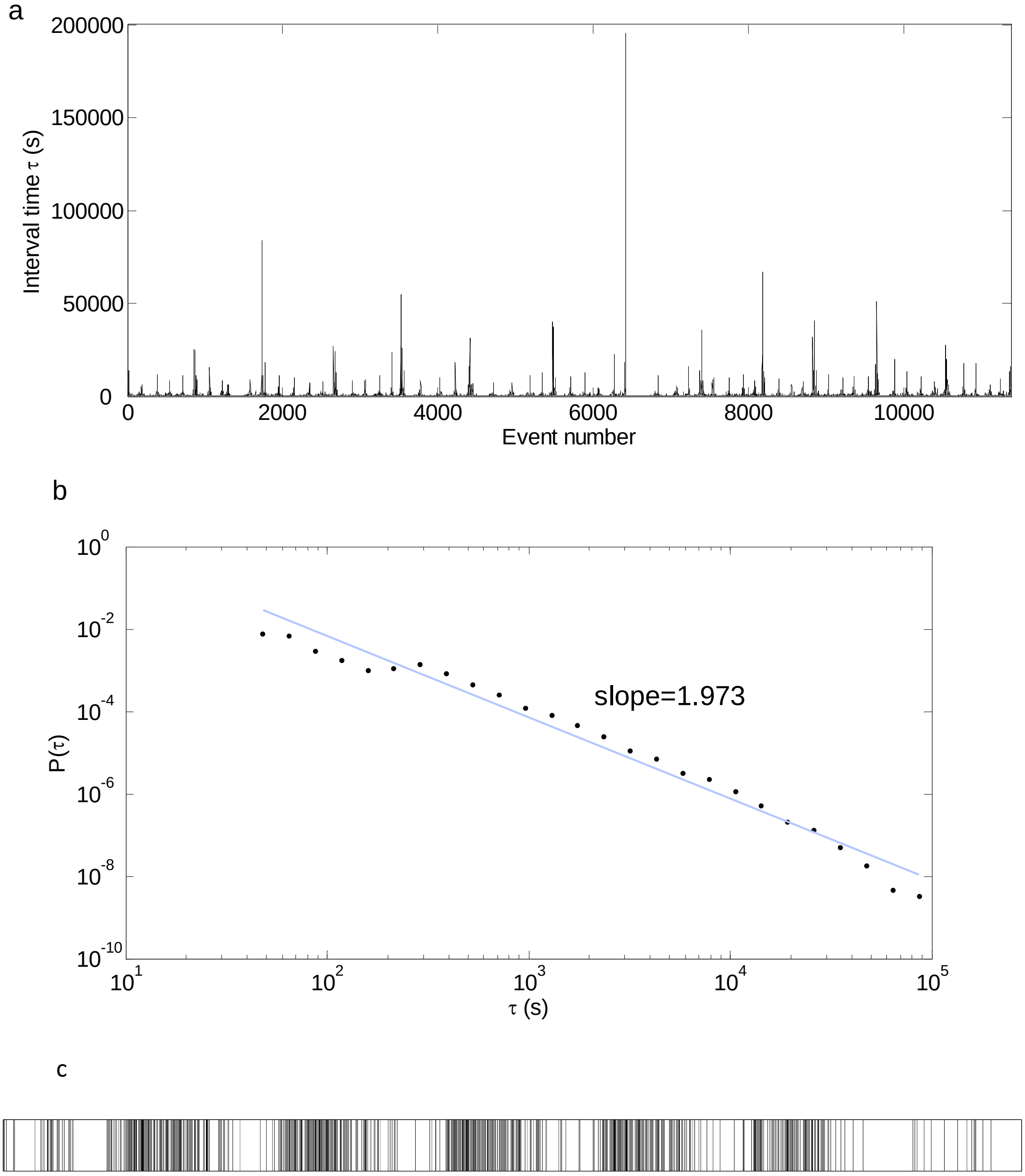

Fig.2 Purchase orders to enterprise A in Q2. **a**, The interval times $\tau$ of the $1.1\times10^4$ consecutive orders in Q2, locating in range from 1s to $2.0\times10^5$ s. Large spikes are visible on this plot as well, although the locations and lengths of these spikes are different from Fig.1 a. **b**, Probability-interval time plot for these orders in doubly logarithmic coordinates, the probability decays in a power-law manner with exponent 1.973. All points can be well fitted by this distribution. **c**, Succession of orders in a certain week of Q2. Five significant bursts and two smaller black boxes can be observed on the plot.

By analyzing the individual characteristics through the PO data in the given two quarters, a heavy-tailed distribution with approximate exponent 2.0 is obtained, although the power-law scales of the two fiscal quarters are a little different. And since the number of purchase orders in a quarter is so large that the succession plot is unreadable, only one week's data are discussed in Fig.1 c and Fig.2 c to provide a snapshot of the succession of orders.

# 4 Group dynamics in the purchasing activity

Although a company assigns PO's to various vendors for different missions, the overall purchasing activity is market oriented under the influence of productivity of the whole industry. Since the market demand varies with time, so the PO assignments in different quarters may show diverse properties accordingly. Without distinguishing any details of these orders in the two quarters, the purchasing characteristic of enterprise M is obtained. On the basis of the fitting results, a model driven by the product life cycle is then introduced.

## 4.1 Statistical characteristics

As described in Sec.3, the heavy-tailed distribution with approximate exponent 2.0 well describes the real feature of purchasing behaviors between enterprise M and a certain supplier. Inspired by the long time inactivity of individual vendor in the purchasing process, we detected the scaling properties of order creation times, including orders to all vendors, aiming to clarify whether the power-law feature is in good agreement with the reality. But to our surprise, the fitting results are not accurate because about one-third of all data points deviate too far from them. Hence, there may be another distribution to better describe the group dynamical features in purchasing activity.

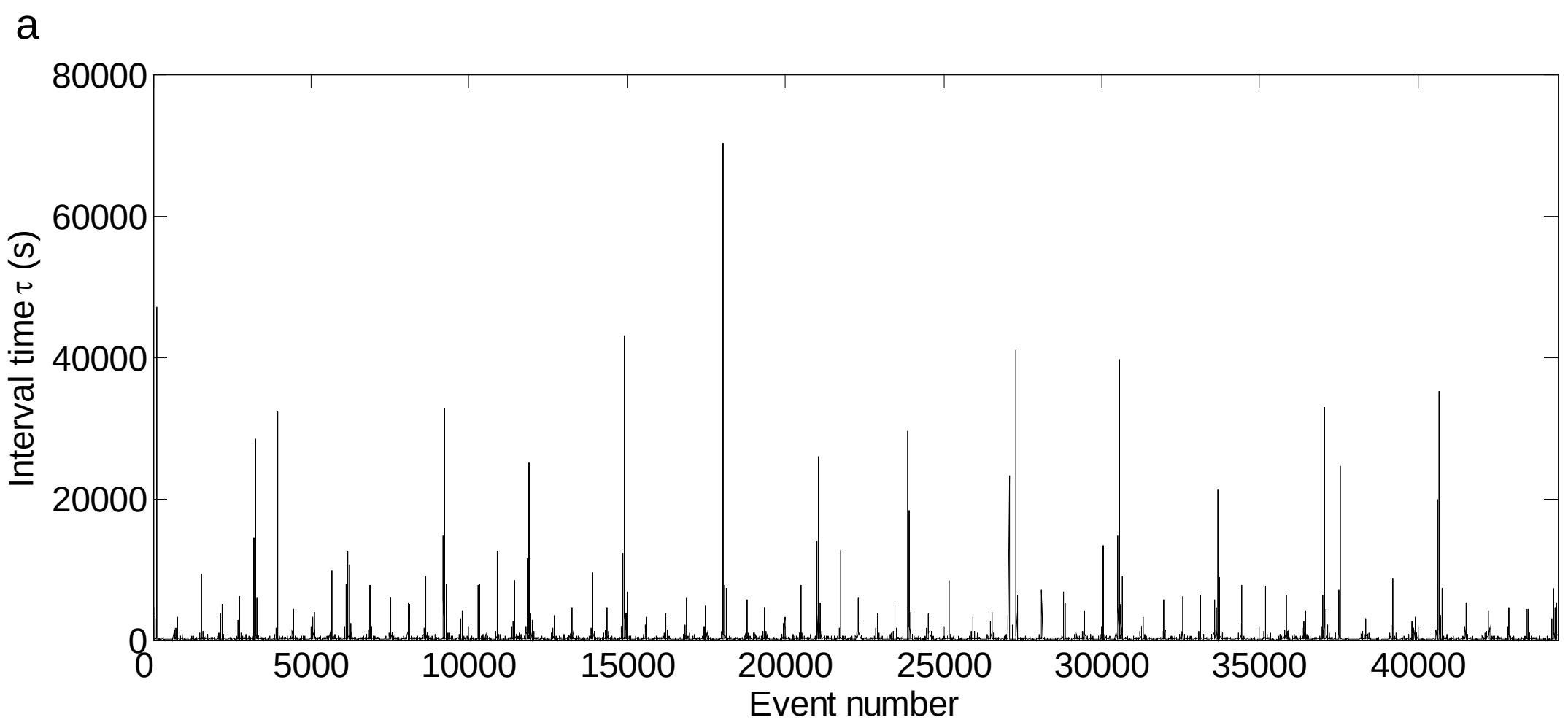

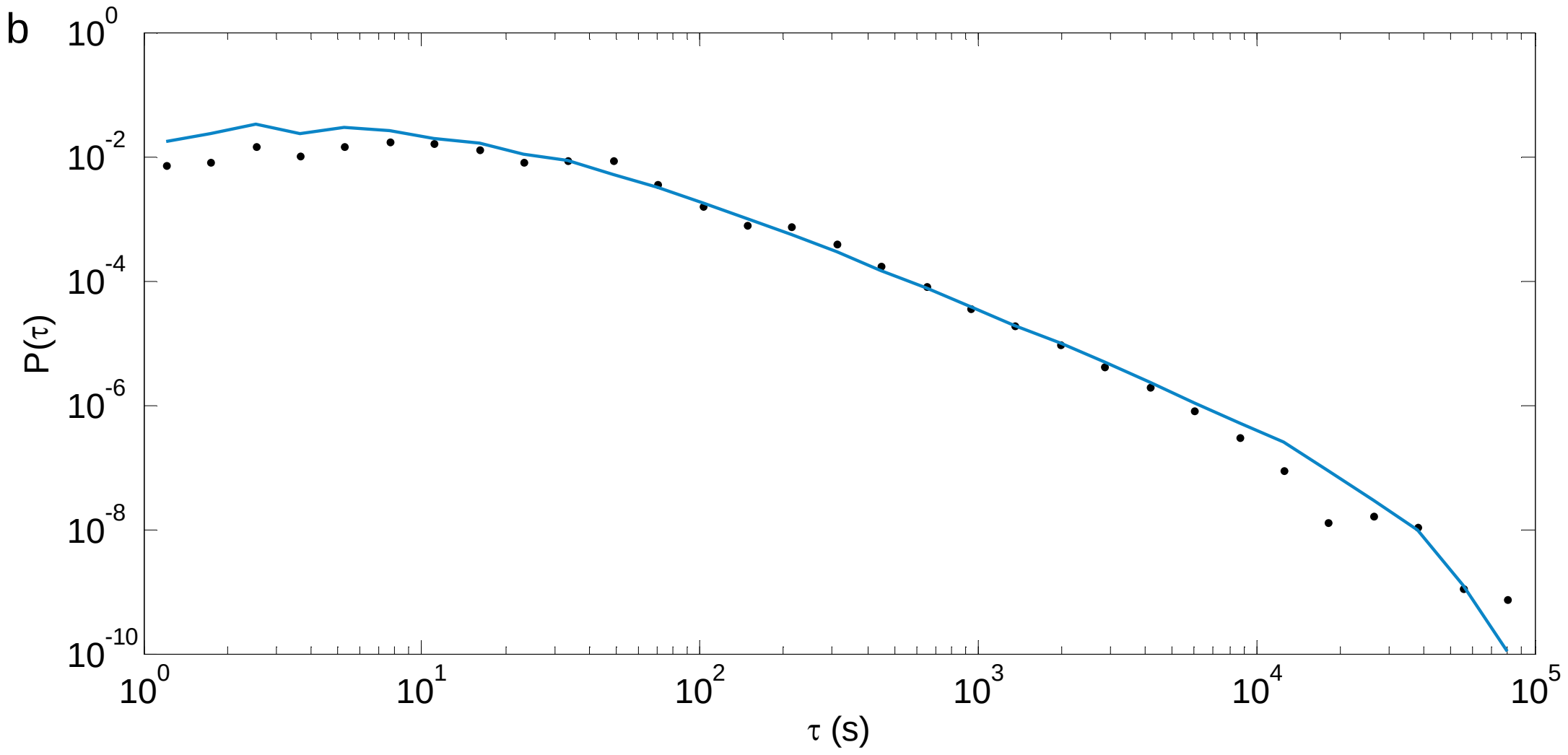

c

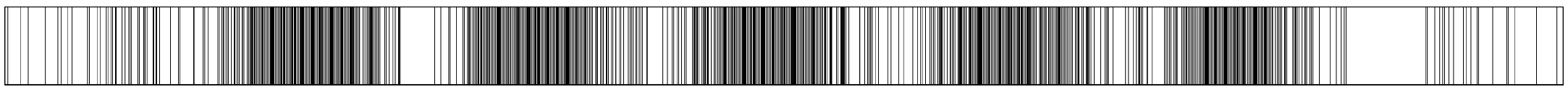

Fig.3 All purchase orders of enterprise M in Q1 without distinguishing the suppliers. **a**, The interval times $\tau$ of the $4.4\times10^{4}$ consecutive PO's in Q1, locating in range between 0s and $7\times10^{4}$ s. Large spikes are still visible. **b**, Probability-interval time plot for these PO's in double-logarithmic coordinates**,** where the probability displays a mixture distribution with both power-law and exponential characteristics $P(\tau) = 20e^{-0.00008\tau}(0.00008(\tau+25)^{-0.9} + 0.9(\tau+25)^{-1.9})$. **c**, Succession of PO's in the same week of Fig.1 c., yet each corresponding vertical line presents the creation of an order. Furthermore, the boxes are bigger and the bursts are much more significant.

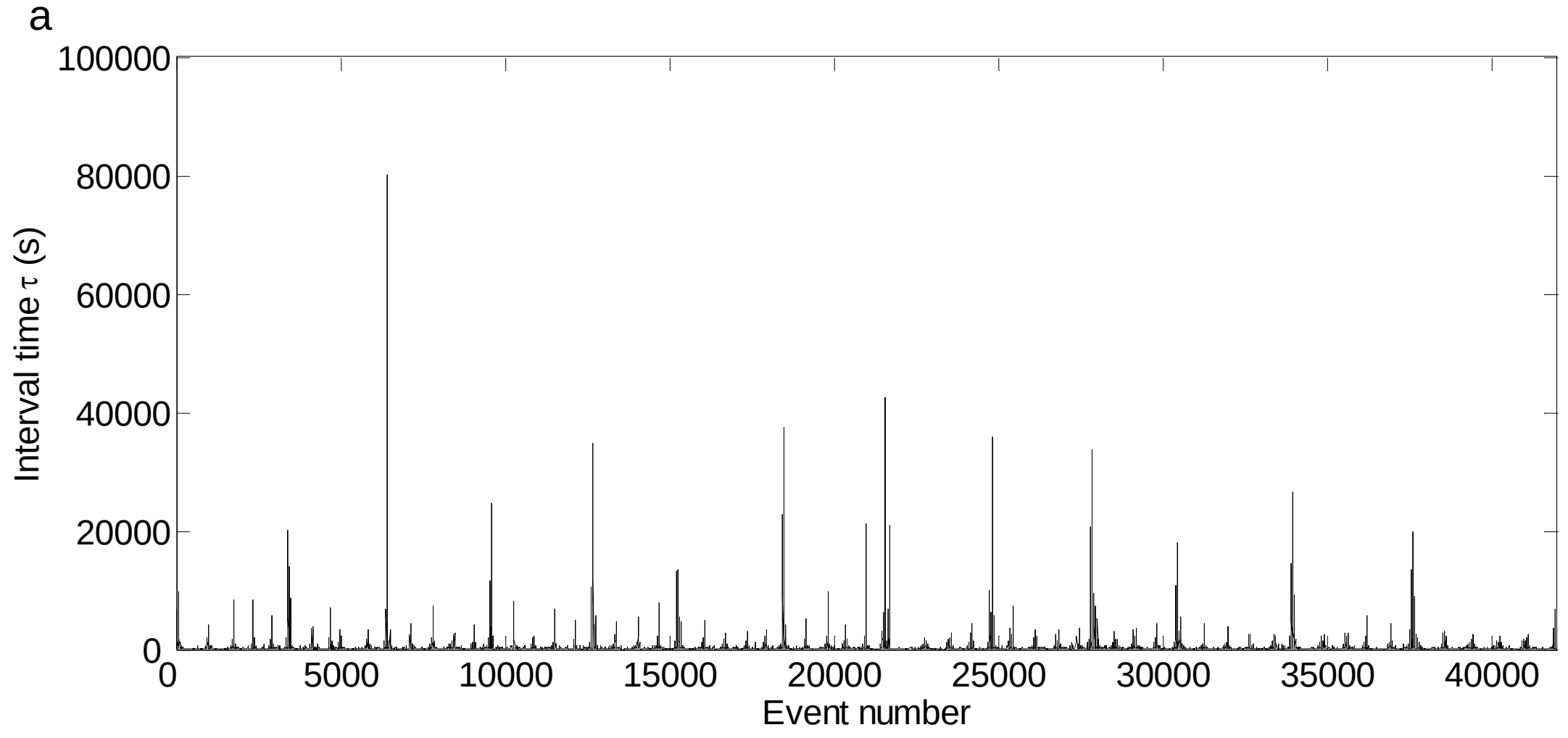

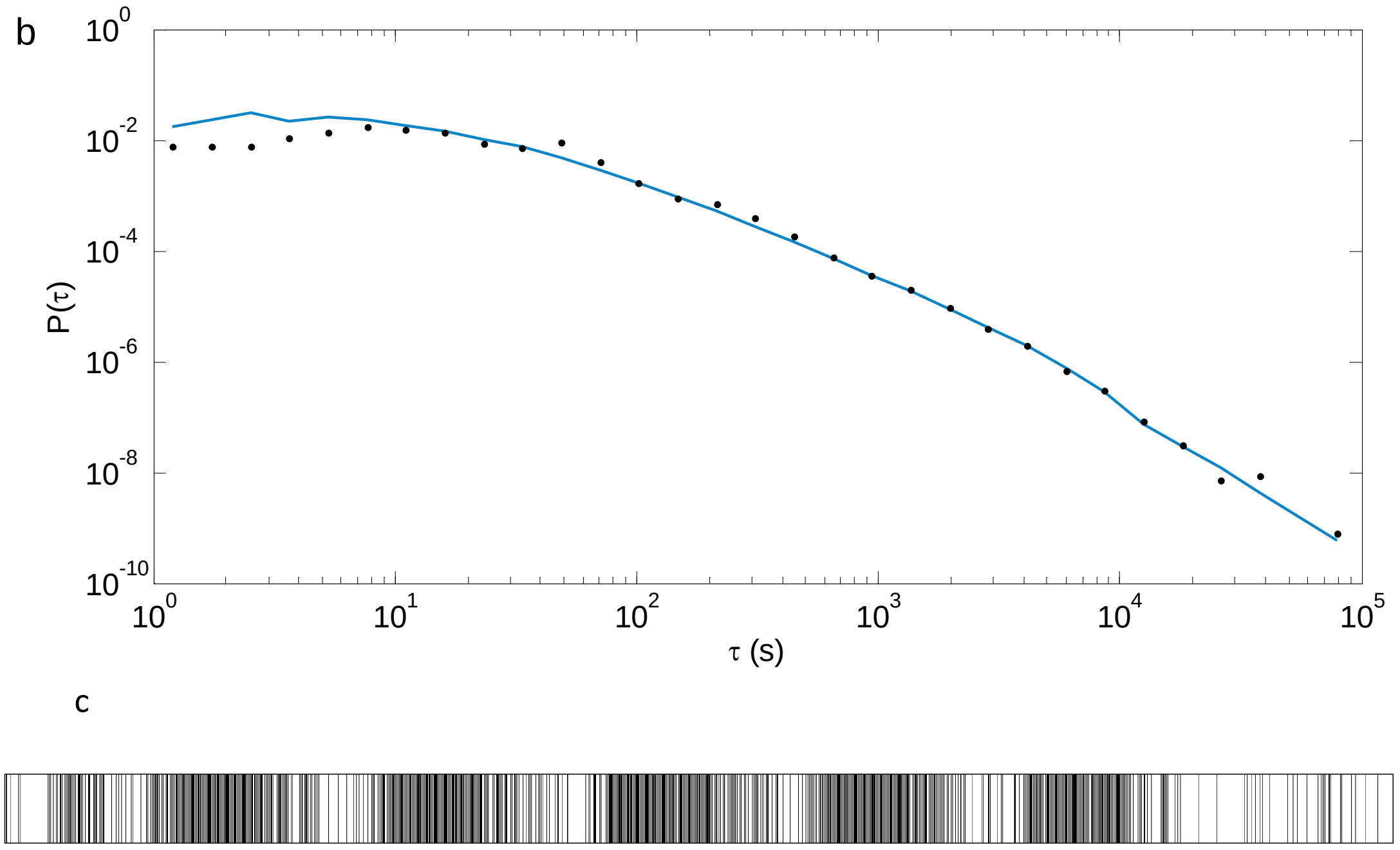

Fig.4 Purchase orders of enterprise M in Q2 without recognition of suppliers. **a**, The interval times $\tau$ of the $4.2\times10^{4}$ consecutive orders in Q2, from 0s to $8.0\times10^{4}$ s. some spikes illustrates the long time of inactivity during this process. **b**, Probability-interval time plot on doubly logarithmic axes, fitting with a mixture distribution $P(\tau)=19e^{-0.0001\tau}\{0.0001(\tau+25)^{-0.9}+0.9(\tau+25)^{-1.9}\}$. **c**, Succession of orders in the same week of Fig.2. Six significant bursts (black boxes on this box) are visible on the plot, although one is significantly smaller than others. In fact, for this week, the number of orders created in Sunday is about one sixth of that on a weekday.

As shown in the above two figures, the arrival process of purchase orders without distinguishing the vendors allows for very long periods of inactivity that separate bursts of intensive activity as well as that on the individual dynamics level. Additionally, the five significant bursts in Fig.3 c illustrate that most of orders are created in weekdays, which means the PO arrival process is affected by working calendar. More precisely, the counts of purchase orders in different weekdays are diverse for the demands are dependent on the stochastic production demands of market in different periods. Furthermore, from Fig.4 c, it is confirmed that orders created in weekends are meaningful and shouldn't be excluded while fitting the regression.

In a word, the mixture distribution is in good line with the reality, although it deviates from the empirical distribution in particular for the first points on the left-hand side of the panel with larger ordinate value. Obviously, this distribution shows significant divergence compared with that in Sec.3. This can be attributed to the fitting deviations as they are negligible for individual dynamics, but those of different vendors are cumulated and lead to an exponential factor for group dynamics.

Considering the real purchasing process in daily operation, we try to introduce a model and analyze the overall purchasing activity from another point of view.

## 4.2 Model analysis

The statistical result implies a fact that a certain dynamical mechanism is in action, which may reveal the overall purchasing process of enterprise M. As a company with fast production update, enterprise M takes a purchasing strategy on the basis of dynamic lot sizing model with lead-time two weeks in response to this short product life cycle. A new PO is placed when the inventory drops to order point or re-order level. The queuing process driven by this strategy can't be explained well by the human dynamics model on task priority. So we try to describe this process with the following model driven by product life cycle:

（i） The frequency of order creation in different time intervals is independent since the requirement is dependent on the stochastic market demand;

（ii） Let the rate of order creation $\lambda(t)=\dfrac{Rt+S_1}{Pt+S}$ at time $t$, where $R, P, S_1, S$ represents demand rate, production rate, initial demand and initial stock respectively;

（iii） The occurring probability of an order at time $t$ is $\lambda(t)dt$, and purchasing behaviors hardly occur more than twice in a short time interval $dt$.

According to the definition above, we obtain $\lambda(t)=\dfrac{R}{P}+\dfrac{PS_1-RS}{P(Pt+S)}$. Hence, based on the approach presented in Ref.15, the cumulative distribution function of the inter-arrival time $P_\tau$ can be given by

$$P_\tau \sim c\left(\frac{P}{S}\right)^{\frac{RS-PS_1}{P^2}}\left(\tau+\frac{S}{P}\right)^{\frac{RS-PS_1}{P^2}} e^{-\frac{R}{P}\tau} \qquad \text{for sufficiently large } \tau$$

Differentiating both sides of the function, we obtain the probability density function

$$P(\tau) \sim c\left(\frac{P}{S}\right)^{\frac{RS-PS_1}{P^2}} e^{-\frac{R}{P}\tau}\left\{\frac{R}{P}\left(\tau+\frac{S}{P}\right)^{\frac{RS-PS_1}{P^2}}+\frac{PS_1-RS}{P^2}\left(\tau+\frac{S}{P}\right)^{\frac{RS-PS_1}{P^2}-1}\right\}$$

where the probability distribution can be approximated by this function in an small interval of length around a specific point $\tau$. As $R, P, S_1$ and $S$ are all positive constants, it denotes that the theoretic inter-arrival time follows a mixture distribution with both exponential and power-law properties.

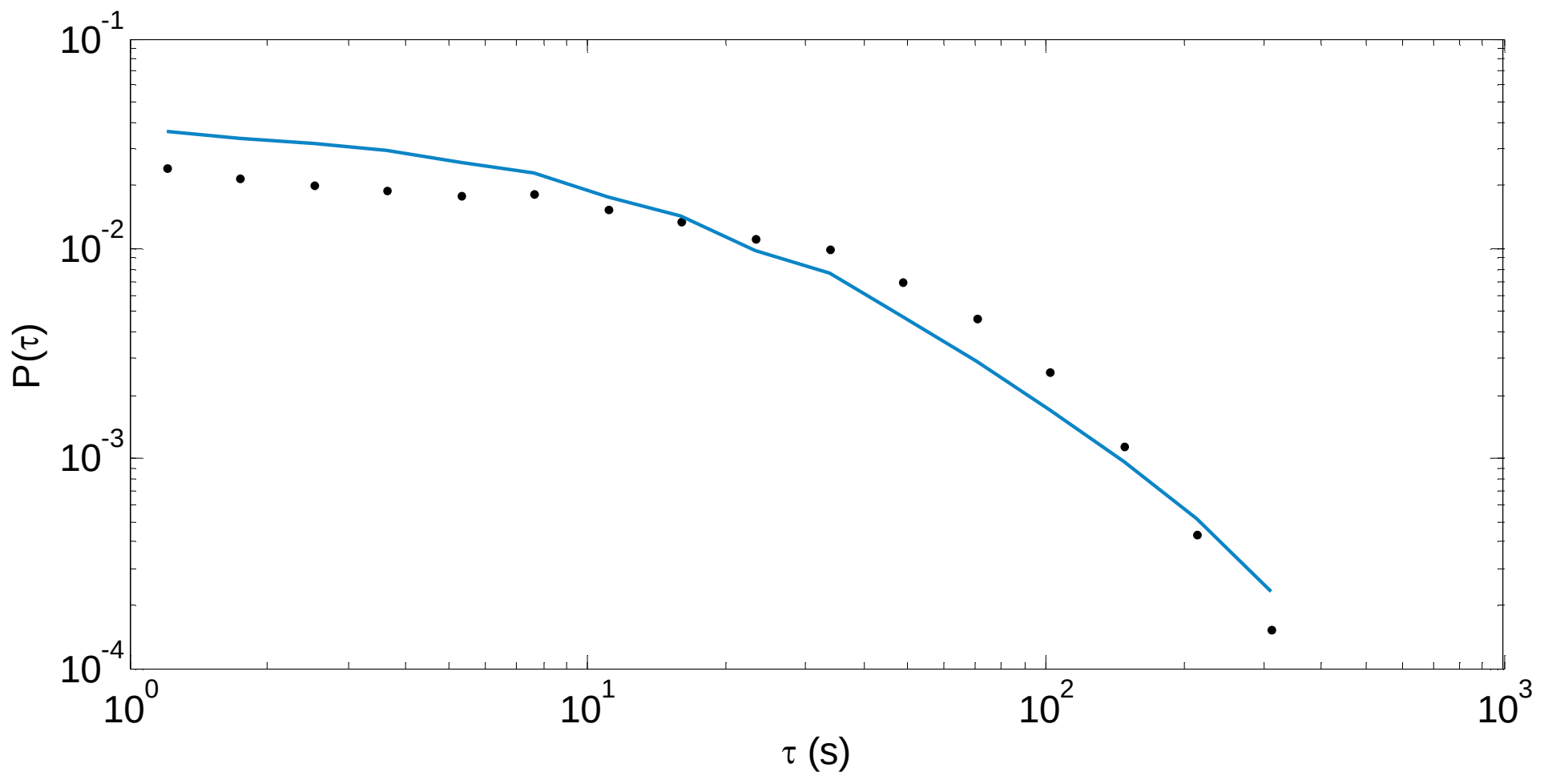

Fig.5 The comparison of the time interval distribution for the activity pattern predicted by the model (black points) and the analytical distribution (blue line) $P(\tau)=19e^{-0.0001\tau}\{0.0001(\tau+25)^{-0.9}+0.9(\tau+25)^{-1.9}\}$ with the same values of $R, P, S_1$ and $S$, where about $10^4$ events was monitored over $T=10^6$ time steps. The simulation result is in good line with the analytical one.

Based on the fitting analysis and the theoretical probability distribution of the model, it is supposed that the overall purchasing activity displays not only a power-law feature but also an exponential characteristic. This verifies that the purchasing strategy plays a significant role in the PO assignment. For both of the two fitting distributions in Fig.3 and Fig.4, the power-law exponents are below zero, by which we can conclude that $\frac{RS-PS_1}{P^2}<0$ in practical operation, i.e. $PS_1-RS>0$. This indicates that purchasing frequency decays over time and tends to a stable value $\frac{R}{P}$ as $t\to\infty$ for order assignment is systematic during the stable production period. While as $t\to 0$, $\lambda(t)$ is larger due to the small order quantity and frequent purchasing actions at the initial stages of product life cycle because the manufacturing process is immature. More specifically, $\lambda(t)=S_1/S$ when $t=0$, and this means that $\lambda(t)$ can be regarded as zero when there is no initial demand or purchasing quantity $S_1$ in the product conception phase.

In short, a heavy-tailed property is observed in both of the two kinds of distributions, although the group dynamical characteristic exhibits an exponential factor besides a power law. This denotes that the purchasing activity is heterogeneous with different characteristics on individual and group dynamics levels respectively.

# 5 Conclusion

This paper investigates whether purchasing behavior has some kind of dynamics by analyzing the PO data and purchasing process. Two distinctive characteristics are found for the interval time distribution from the point of view of individual dynamics and group dynamics. The decay characteristics of purchasing activity from enterprise M to supplier A are likely power-law with approximate exponent 2.0 applying the log binning method, although the scaling parameters vary in different quarters. However, for the characteristic of overall purchasing activity, the interval time distribution displays both power-law feature and exponential property. The observation of heavy-tailed in the statistical data reveals a typical purchasing activity pattern in the protean industrial environment with fast production update. Indeed, the purchasing frequency depends on the specific demand rate, production rate under the conditions of competitive market and short production life cycle. The model can address the scaling behavior of the probability distribution with the analytical function and the simulation result.

Understanding the origin of the non-Poisson nature of purchasing activity pattern is meaningful for effective resource allocation and better performance of supply chain management. However, considering the complicated environment for diverse industries, one may raise several important questions. Does the power-law feature exist in purchasing behaviors of other companies? May the model be applied to other human actions and are there any models from other distinct perspectives to interpret this group dynamical characteristic? Why the characteristics on the group dynamics and individual dynamics are diverse? In the future, more data of different corporations need to be collected and much exacter model will be developed to offer a better understanding of human dynamics and supply chain management.

## Acknowledgments

This paper is partly supported by the National Natural Science Foundation of China (70871082), the Foundation of Shanghai Leading Academic Discipline Project (S30504) and the Innovation Fund Project for Graduate Student of Shanghai (JWCXSL1002).

## Reference

[1] A.-L. Barabási, The origin of bursts and heavy-tails in human dynamics, Nature 435 (2005) 207-211.
[2] J.G. Oliveira, A.-L. Barabási, Darwin and Einstein correspondence patterns, Nature 437(2005) 1251.
[3] A.Vázquez, Exact Results for the Barabási Model of Human Dynamics, Phys. Rev. Lett. 95(2005) 248701.
[4] Z. Dezsö, A. Lukács, B. Rácz, I. Szakadát, A.-L. Barabási, Dynamics of information access on the web, Phys. Rev. E 73 (2006) 066132.
[5] W. Hong, X.-P. Han, T. Zhou, B.-H.Wang, Heavy-tailed statistics in short-message communication, Chinese Physics Letters 26 (2009) 028902.

[6] D. Brockmann, L. Hufnagel, T. Geisel, The scaling laws of human travel, Nature 439(2006) 462-465.

[7] Bagler G. Analysis of the Airport Network of India as a complex weighted network. Physica A 387 (2008) 2972—2980.

[8] A.Vázquez, J. G. Oliveira, Z. Dezsö, K. Goh, I. Kondor, A.-L. Barabási, Modeling bursts and heavy tails in human dynamics, Phys. Rev. E 73(2006) 036127.

[9] Q. Wang, J.-L. Guo, Human dynamics scaling characteristics for aerial inbound logistics operation, Physica A 389 (2010) 2127-2133.

[10] G. Caldarelli, S. Battiston, D. Garlaschelli, M. Catanzaro, Emergence of Complexity in Financial Networks, Lect. Notes Phys. 650(2004) 399-423.

[11] D.-S. Ho, C.-K. Lee, C.-C. Wang, M. Chuang, Scaling characteristics in the Taiwan stock market, Physica A 332 (2004) 448-460.

[12] H.-J. Sun, J.-J. Wu, Scale-Free Characteristics of Supply Chain Distribution Networks, Modern Physics Letters B 19 (2005) 841-848.

[13] E.P. White, B.J. Enquist, J.L. Green, On Estimating the exponent of Power-law Frequency Distributions, Ecology 89(2008) 905-912.

[14] M.E.J. Newman, Power laws, Pareto distributions and Zipf's law, Contemporary Physics 46(2005) 323-351.

[15] J.-L.Guo, Weblog patterns and human dynamics with decreasing interest, arXiv:1008.0042v3 (2010).